# Effect of a Cone Shape on the Motion of Active Janus Colloids

Zhiyuan Zhao[1,2,3], Teun W.J. Verouden[2,3], Dillip K. Mohapatra[2,3], Evert M.E. Simons[2,3], Janne-Mieke Meijer[2,3,a)]

AFFILIATIONS

[1]Green and Low-Carbon College, Yancheng Teachers University, Yancheng, 224000, China

[2]Department of Applied Physics and Science Education, Eindhoven University of Technology, Groene Loper 19, 5612 AP Eindhoven, The Netherlands

[3]Institute for Complex Molecular Systems, Eindhoven University of Technology, Groene Loper 19, 5612 AP Eindhoven, The Netherlands

[a)]Author to whom correspondence should be addressed: j.m.meijer@tue.nl

**Abstract**

The propulsion of active colloids is governed by their symmetry and shape, yet a systematic investigation of how small geometric effects influence the locomotion of anisotropic active colloids is lacking. In this paper, we study the effect of a cone shape by combining high-resolution two-photon polymerization 3D printing with AC electric field experiments to study anisotropic Janus micro swimmers. We fabricate a series of Janus colloids with different shapes, ranging from a sphere to a hemisphere with a cone-protrusion, but containing similar hemispherical gold-coatings. Under an applied alternating current (AC) electric field, all particles exhibit active, ballistic motion. We find a shape and size dependence on the propulsion velocity, with the spherical Janus particle moving fastest, followed by a small cone and a large cone protrusion, while surprisingly a printed sphere with a flat side moves slowest. In addition, a reversal in the direction of motion for all shapes was triggered, a phenomenon governed by a transition from induced-charge electrophoresis (ICEP) at low frequencies to self-dielectrophoresis (sDEP) at high frequencies. We reveal a distinct shape influence, with the large cone-protrusion increasing their velocity in the sDEP regime. Our results provide insight into the link between active particle geometry and their propulsion velocity that are important for understanding biological micro swimmers and designing optimized microrobots.

## 1. Introduction

Active matter, comprised of vast numbers of individual "agents," is ubiquitous in nature. These agents consume energy to move or generate mechanical forces, as seen in phenomena like bacterial swarming, bird flocking, and fish schooling[1-3]. A key class of active matter are synthetic colloidal micro swimmers, also known as active colloids. Active colloids serve as model systems for testing theoretical predictions and uncovering insights into the locomotion of biological microswimmers[4,5]. At the same time, active colloids have become of increasing interest for medical and adaptable material applications, such as cargo delivery[6], sensing[5], and microrobotics[7].

The self-propulsion of active colloids is achieved by breaking the symmetry of the force distribution on the particle's surface. In a continuous medium, this symmetry breaking can be accomplished by engineering the particle's surface properties[8,9]. A typical example are microspheres of which one hemisphere is coated by gold, known as Janus colloids, that exhibit superdiffusive behavior upon conversion of hydrogen peroxide at the gold surface resulting in a chemical gradient[10]. Besides chemical gradients, many other external energy sources can be employed to induce active motion of Janus colloids, including magnetic[11], acoustic[12], optical[13], and AC electric[14] fields. Of these methods, AC electric fields are particularly attractive as they provide a facile, remote means of powering active particles without chemical fuel consumption, i.e. active motion is present as long as the electric field is applied. The active motion is the result of induced-charge electrophoresis (ICEP), driven by unbalanced electro-osmotic flows around the dielectric and conductive sections of metal-coated particles[9]. Consequently, the active motion can be finely tuned via the AC signal parameters[15]. By controlling the applied frequency the active Janus colloid motion can be controlled forward or backward with respect to their metal-coated side which is attributed to self-dielectrophoresis (sDEP [16].

Thanks to advances in chemical and lithographic techniques, the study of active colloids with irregular and complex particles shapes has become possible and revealed that symmetry and shape matter[17]. Several different anisotropic shapes have been studied, including L-shaped[18], helical[19], and ellipsoidal particles[20], as well as microspinners[21] and patchy particles[22], and were found to show much more complex propulsion modes, such as spinning, steering, and braking[1]. An approach that has immense potential to study the motion of anisotropic swimmers is 3D nano printing. 3D nano printing offers significant flexibility in shape and symmetry of structures with sub-micron resolution[23,24]. The method facilitates the fabrication of colloidal particles with a wide range of geometries with gradual shape tuning and in sufficient quantities for systematic studies. For instance, 3D printing has been used to fabricate colloidal particles with complex surface topologies to study hierarchical self-assembly[25]. Preparing anisotropic active colloids via 3D printing is uniquely powerful. For instance, the specific tapered design of the propulsion side can increase the swimming speed significantly[26]. Other studies revealed a clear shape dependent behavior for helical shapes[27] and cluster formation when interlocking particles are employed[28]. However, a systematic comparison between studies of different active anisotropic particle shapes remains difficult, as often between studies different propulsion mechanisms are employed or changes in particle shape coincide with changes in the size and shape of the symmetry breaking component, e.g. metal-coating, which for spherical particles is already known to change the particle behavior[15].

Here, we employ two-photon polymerization 3D printing to fabricate active anisotropic Janus particles with shapes derived from a sphere and investigate their self-propelled behavior under AC electric fields. To systematically study the influence of a gradual geometric change, we prepared a series of colloids with different shapes while keeping the metal coated side similar in size and shape, i.e. a hemisphere. The particle shapes range from a sphere, a hemisphere with a small cone, and a hemisphere with a long cone. In addition, we compare the motion to a spherical colloid of similar size. To study how symmetry and shape couple to motility and motion trajectories, we expose all Janus colloids particles to the same AC electric fields. We observe and track the particle trajectories with video microscopy to measure and compare the velocities for the different shaped particles under the same propulsion conditions. In addition, we investigated the particle trajectories across a wide frequency range, to see if a reversal in their direction of motion occurs due to ICEP and sDEP propulsion mechanisms. Our results provide insight into how the exact geometrical shape and size of an active colloid influences their propulsion.

## 2. Materials and Methods

*2.1 Fabrication of active anisotropic particles*

To prepare spherical Janus colloids, a polystyrene sphere (3 μm, Sigma-Aldrich) monolayer was prepared by a gas/liquid interface method[29]. Next, the sample was placed into a sputtering coater (Quorum Technologies, Q150TS) and a 5 nm Cr and 30 nm Au layer were deposited. After the deposition, the sample was immersed in deionized water to disperse the Janus spheres by mild ultra-sonication, and then centrifugally cleaned with deionized water. A schematic illustration of the process is given in Figure 1a. To study the influence of a cone shape, three distinct 3D particle shapes; a sphere, a hemisphere with small truncated-cone (HS-Cone2), and a hemisphere with long truncated-cone (HS-Cone4) were designed by 3D Max as schematically shown in Figure 1(b-d). To allow comparison with spherical Janus colloids, we employed the same diameter for the printed sphere and hemisphere of 3 μm. The small cone has a height of 2 μm while the long cone has a height of 4 μm and both have a base diameter of 2 μm. Next, these designs were exported to the standard tessellation language (STL) file and processed by DeScribe, which is used to tile the STL file in a small repeating array (20 × 20), which is part of a larger repeating array (15 × 15). Particles were printed on glass coverslips (22 mm in diameter) with a thickness of 0.16 mm in the photoresist IP-Dip2 (Nanoscribe GmbH & Co.KG) using a Nanoscribe Quantum X Shape equipped with a 63 × 1.40 NA objective. Afterwards, the sample was developed in propylene glycol monomethyl ether acetate (PGMA, >= 99.5%, Sigma-Aldrich) for 30 min and subsequently in 2-propanol (>= 99.5%, Sigma-Aldrich) for 5 min. After drying, the particles were coated with a 5 nm layer of Cr and a 30 nm layer of Au using a sputtering coater. After sputter coating, the glass coverslip is cut to fit in a 1.5 mL centrifuge tube and submerged in 1 wt% Pluronic F127 solution for 5 min. The sample is then removed from the centrifuge tube and resubmerged in a 1.5 mL centrifuge tube containing deionized water. The centrifuge tube is sonicated for 1 min to release the particles from the substrate. Particles were subsequently concentrated by centrifugation followed by removal of the supernatant.

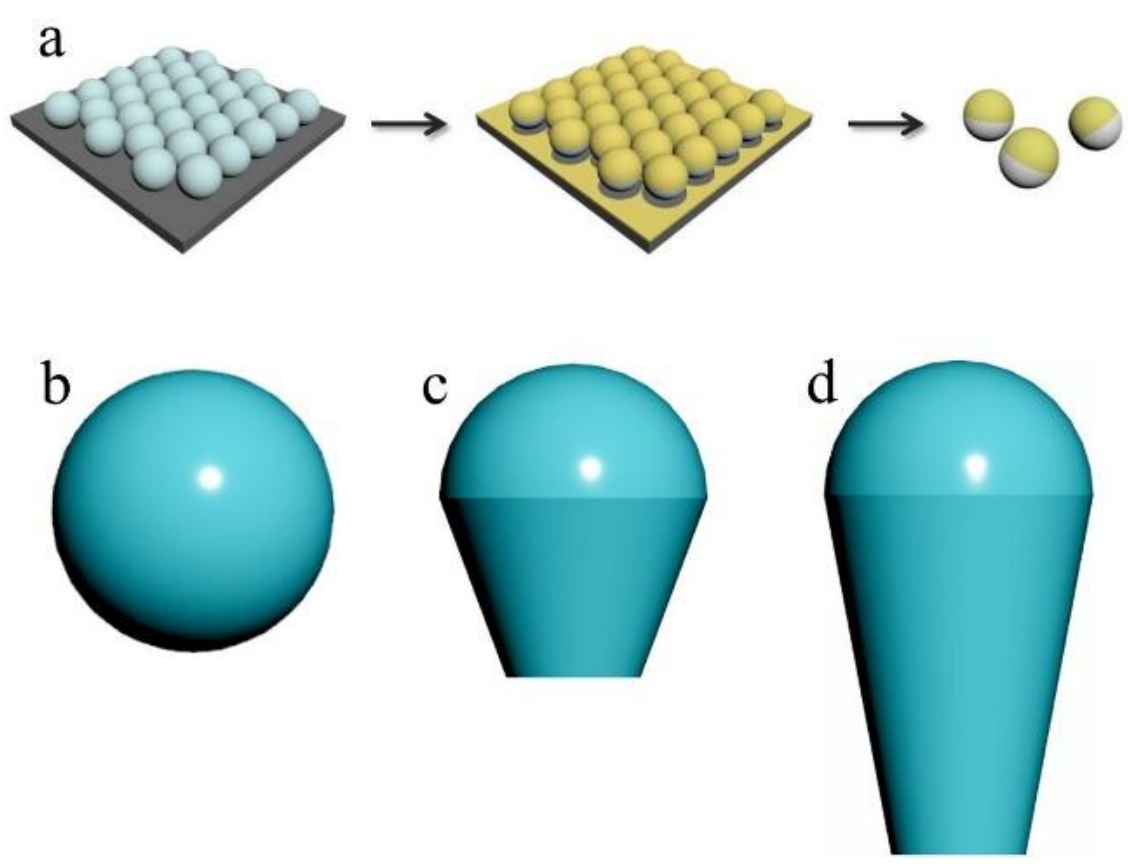

**Figure 1.** (a) Illustration of fabricating Janus spheres half coated with 5-nm chromium and 30-nm gold. After metal deposition the Janus spheres were dispersed in DI-water by mild ultra-sonication. (b-d) Schematics of 3D printed particle designs (b) a sphere, (c) a hemisphere with small cone (HS-Cone2), (d) a hemisphere with long cone (HS-Cone4).

*2.2 Particle Characterization*

Scanning Electron Microscopy (SEM) images were taken with a Thermo-Fisher Quanta 200 3DFEG. Samples consisted of either an array of particles coated with 10 nm gold after printing, or particles after coating and removal from the substrate, in which case the resulting particle solution was deposited onto a silicon wafer and allowed to dry in air.

*2.3 Video Microscopy*

Electric field cells were constructed in-house from 20 x 20 mm ITO coated coverslips (Diamond Coatings LTD, 8-12 Ohms, No. 1.5) (Figure 2). The particle suspension was sandwiched between two conductive transparent ITO coated coverslips and subjected to an AC electric field in the vertical direction. The electrodes were separated by a 120-μm thick spacer with a 51-mm aperture (SecureSeal Imaging). After waiting for a couple of minutes, due to their size all the Janus colloids sedimented to form a quasi-two-dimensional layer on the bottom electrode. By applying an AC electric field of 6 $V_{pp}$, the Janus particles started to move around in this quasi-two-dimensional plane, which is perpendicular to the electric field. The particle motion was imaged using a Nikon Ti2 inverted microscope in Bright Field mode equipped with a 40× objective and a CMOS camera. Movies were recorded at the rate of 10 frames per second with an exposure time of 10 ms.

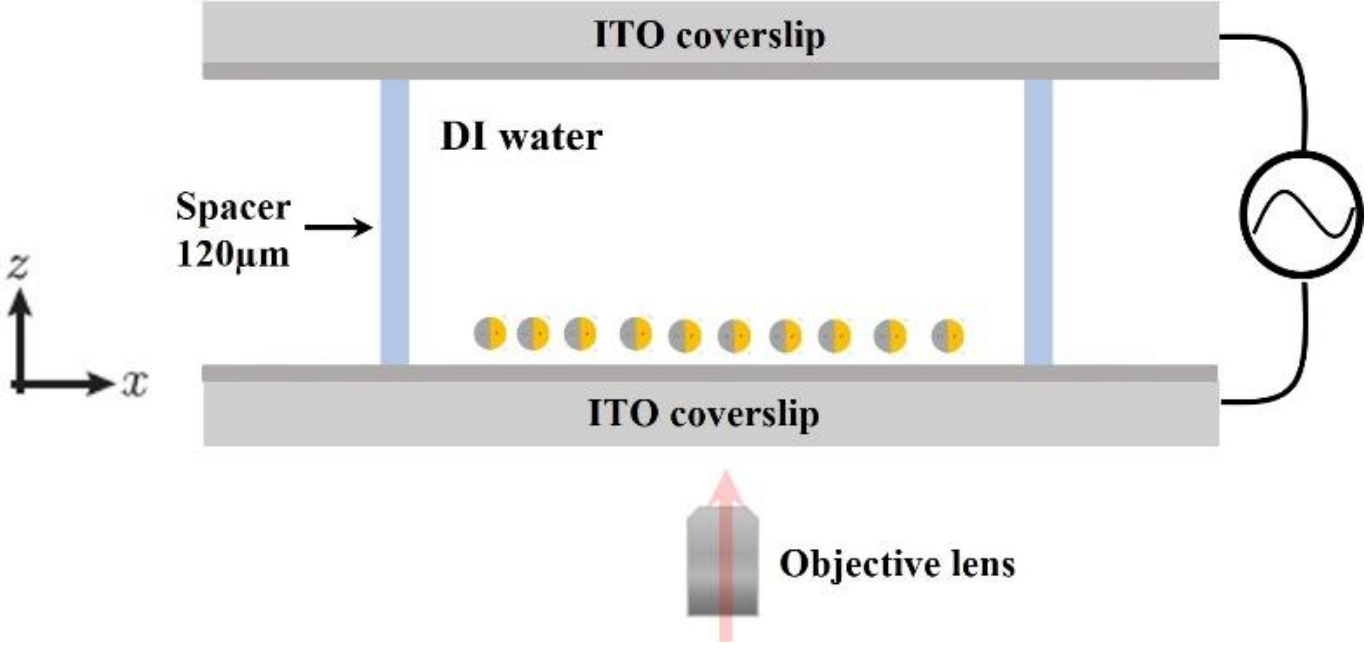

**Figure 2.** Schematic representation of the electric field sample cell employed for studying the dispersions of Janus colloids.

*2.4 Data analysis*

Particle trajectories were extracted from the video data using the Trackpy algorithm[30], a Python-based implementation of the Crocker and Grier method[31]. From these trajectories, the mean-squared displacement (MSD) was calculated for each particle. The translational velocities were subsequently derived from the MSD profiles, following an established analytical framework for quasi-two-dimensional systems[32]. The particle velocity, $v$, of the active particles was determined by fitting the MSD data in the active regime to the equation MSD $= \langle \Delta r^2 \rangle = 4Dt + v^2t^2$, where $t$ represents the lag time and $D$ the translational diffusion coefficient. The fitting was performed using a standard least-squares regression method.

## 3. Results and discussion

*3.1 Fabrication of active anisotropic particles*

The advantages of using 3D printing make it very simple to build complex anisotropic particles, such as stars, letters, numbers, polyhedrons, and polypods. We utilized two-photon polymerization to 3D print three shapes of colloidal particles, including a printed sphere, HS-Cone2, and HS-Cone4. The arrays of the printed particles were investigated with SEM and are shown in Figure 3. We note that the distance between the HS-Cone4 particles is 10 μm to avoid particle tipping due to the action of capillary force (which was observed at closer particle spacings). From the SEM images we can see that the printed sphere appears round at the top but truncated at the bottom, indicating that printing perfect 3 μm spheres was not achieved. For both cone-shaped particles the SEM images confirm that a hemisphere with a cone shape was achieved.

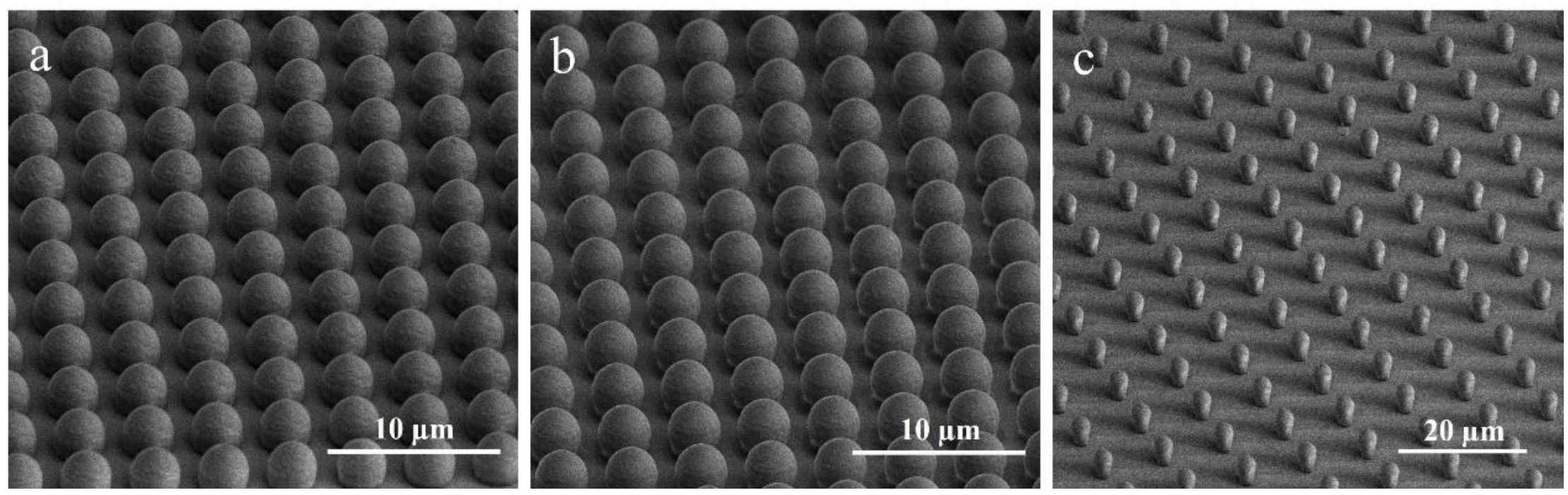

**Figure 3.** SEM images of the arrays of the printed particles taken at a 45° tilt. (a) printed sphere, (b) HS-Cone2, (c) HS-Cone4.

*3.2 Electric field induced motion*

To render the particles active in an AC electric field and turn these into Janus colloids, the top hemisphere of the printed spheres and the two hemisphere-cone shaped particles were coated with a chromium and gold layer. After coating, the particles were transferred from the substrate into an aqueous medium. SEM images of all four particles after coating are presented in Figure 4a1-d1, where the brighter region on one side of each particle corresponds to the gold coating. Due to the inherent nature of the additive manufacturing process, the printed particles exhibit a minor degree of surface roughness. This roughness can be mitigated by minimizing the inter-layer distance during printing. It is also noteworthy that the printed spheres possess a flat base, which is an artifact of the particle-substrate interface. This feature arises from the elongated shape of the printing voxel along the axis of the writing beam and a slight embedding of the print into the substrate, which is necessary to ensure adhesion and stability during fabrication. However, overall, for all particles we find that the gold coated hemisphere appears very similar between all particles, while the non-coated side are distinctly differently shaped. Therefore, this set of particle shapes does allow us to investigate the effect of a cone shape on the active particle motion in an AC electric field.

Particle shape is a critical factor governing hydrodynamics and interparticle interactions. Specifically, the motion and self-assembly capabilities of nano- and microparticles are predominantly dictated by their geometry[33]. After being printed and released into water, the dynamics of the 3D-printed colloidal particles were studied using bright-field microscopy. A uniform AC electric field was applied to induce self-propulsive behavior in these metallodielectric anisotropic particles [16] and their particle trajectories extracted. As illustrated in Figure 4a2-d2, the fabricated active particles exhibited swimming motion under the AC electric field. We note that when the electric field is turned off, the active motion ceases, and the particles diffuse freely in the dispersion medium. Multiple trajectories of individual particles are overlaid in Figure 4a3-d3, with distinct colors representing different particle paths. Clear differences can be observed between the different particle shapes. For both the spherical and the hemisphere-Cone2 the particles cover a greater distance, and the motion is considerably more directed than expected for particles moving solely due to Brownian motion but still shows some rotational diffusion. For the printed sphere and the hemisphere-Cone4 we observe shorter trajectories with a relatively straight path. Clearly, since the gold hemisphere is almost identical, it must be the particle shape and size that leads to the different active motion of these Janus colloids.

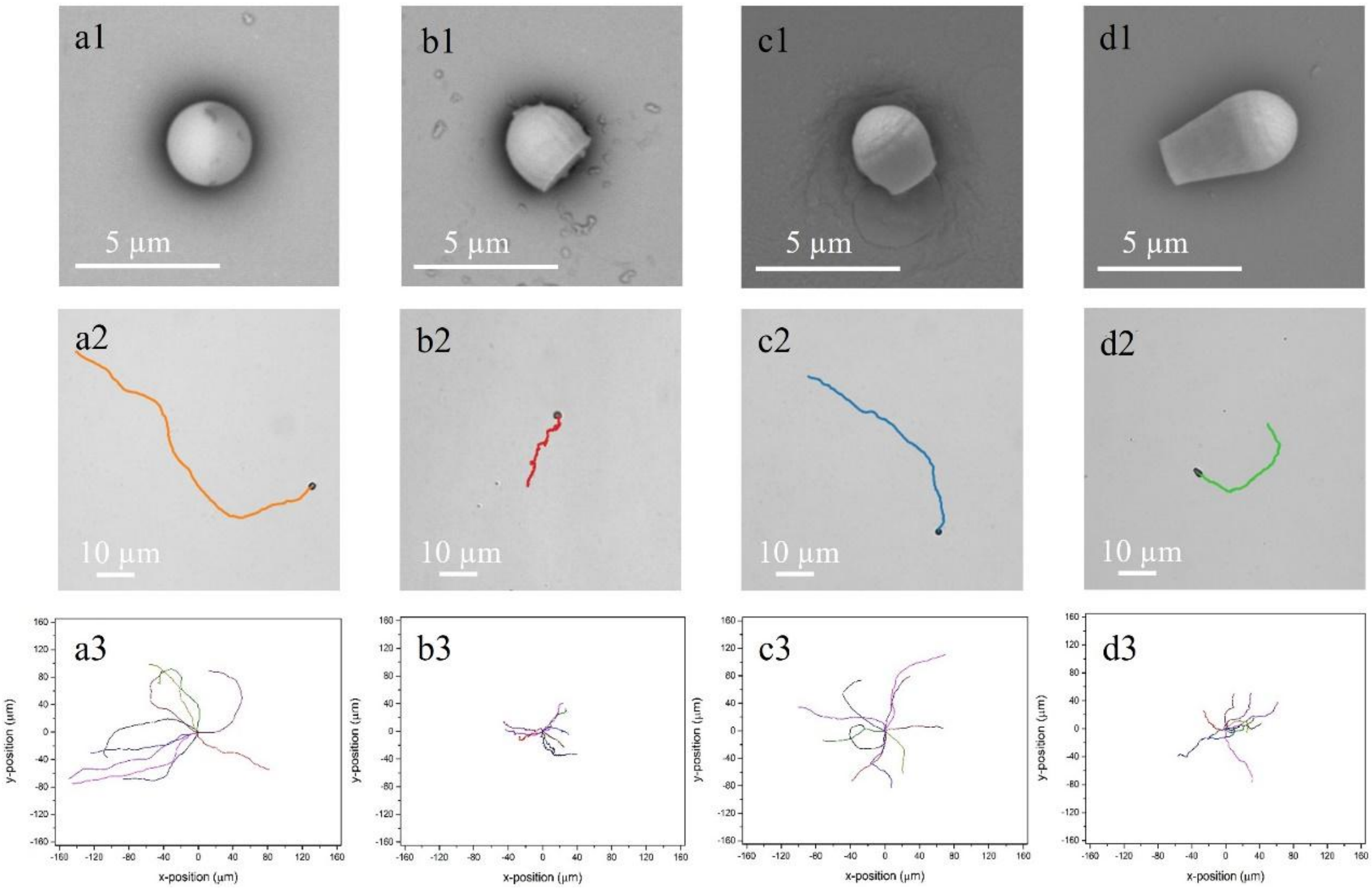

**Figure 4.** Motion behavior of the four shapes of particles (a) sphere, (b) printed sphere, (c) hemisphere-cone2 and (d) hemisphere-cone4. (a1) - (d1) SEM images of different shaped individual particles after sputter-coating with Au and removal from the substrate. (a2) - (d2) Examples of 30 s trajectories showing the active behavior of sphere, printed sphere, HS-Cone2, and HS-Cone4 under AC electric field (voltage is 6 $V_{pp}$, frequency is 10 kHz). (a3) - (d3) Multiple trajectories of individual particles centered at the origin recorded over the course of 30 s.

To quantify the effect of shape, we determined the activity of each particle shape by measuring the MSD over time. We calculated the MSD curves from the trajectory of all active particles observed in 30s video with a frame rate of 10 fps. Figure 5 presents the resulting MSD curves for the four particle shapes at a driving frequency of 10 kHz, all exhibiting characteristics typical of an active colloidal system. The MSD curves were fitted with $\langle \Delta r^2 \rangle = A \Delta t^n$ , where n is the exponent. For active particles ballistic motion captured by $n = 2$ is expected. The measured exponents for our particles ranged from 1.8 to 1.95, confirming their active, self-propelling nature. Furthermore, the average velocities of the microswimmers were calculated from the MSD data. The average propulsion velocity, $v$, of the particles was determined and found for the spheres $v_S = 8.9 \pm 0.1$ μm/s, for the printed spheres $v = 2.6 \pm 0.1$ μm/s, for HS-cone2 $v = 7.4 \pm 0.1$ μm/s, and for HS-Cone4 $v = 4.0 \pm 0.1$ μm/s. Clearly, the propulsion velocity decreases with the increasing conical deformation of the particles. We can partly explain this by the increased weight of the particles, e.g. the volume of the HS-Cone4 is almost twice that of the other particles. Surprisingly, however, we find that the velocity of the printed spheres is the slowest, which seems to be caused by the presence of a flat side, where the sphere was connected to the substrate surface (see Fig. 4b1).

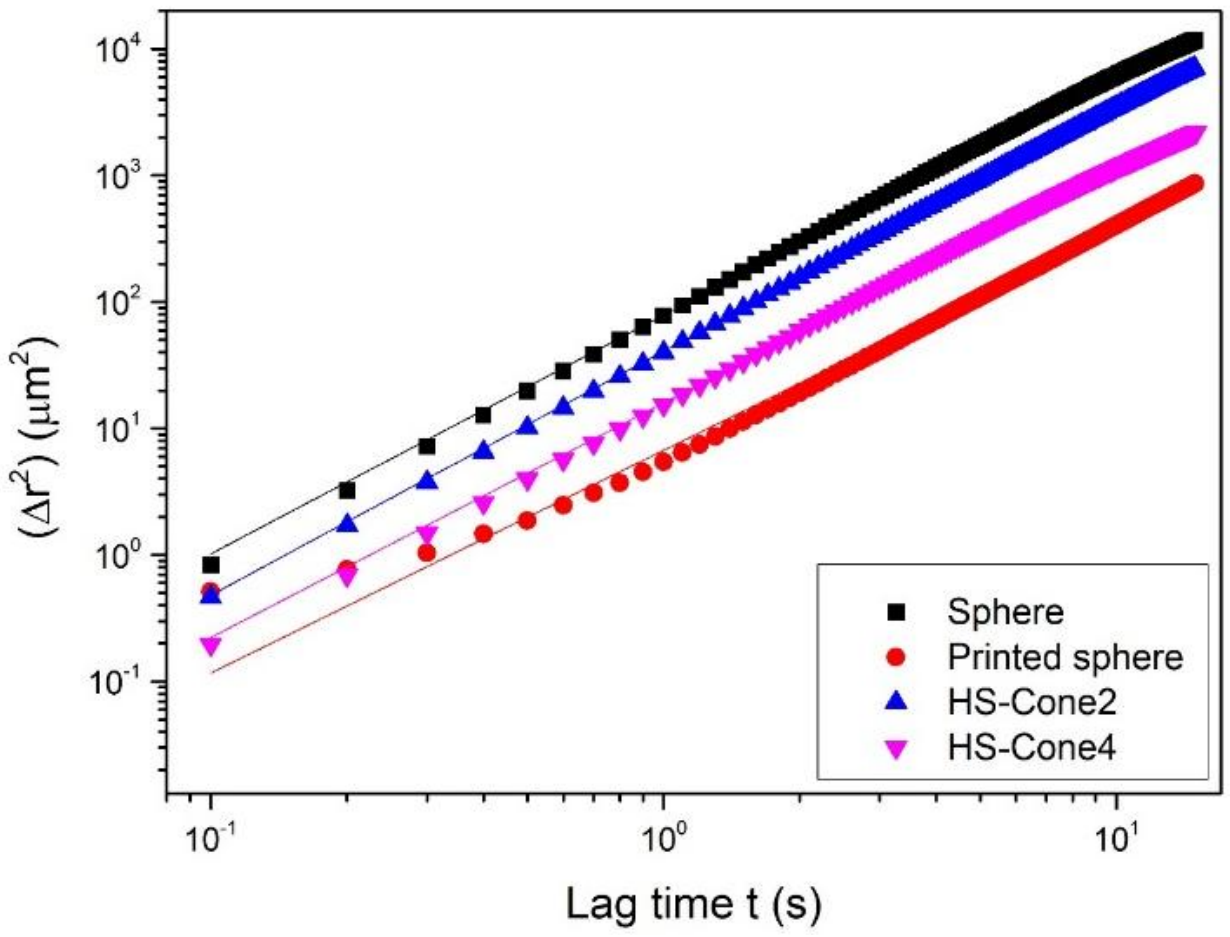

**Figure 5**. Mean squared displacements profiles of four shapes of active particles are shown in log-log plot. The applied voltage is 6 $V_{pp}$ at a frequency of 10 kHz.

The propulsion of metallodielectric Janus colloids induced by self-induced field gradients, is frequency dependent[16]. At low frequencies (typically below 50 kHz), Janus particles move towards their dielectric hemispheres, due to ICEP. However, above a distinct threshold frequency, the particles will undergo a reversal in their direction of motion, propelling with their metallic caps leading. This reversal can be attributed to sDEP[16]. In the high-frequency regime, the sDEP force arises from the interaction between localized, non-uniform electric field gradients and the induced dipole in the metallic coating. In both the ICEP and sDEP regimes, the particle's orientation is governed by the alignment of its induced dipole with the applied electric field. This alignment maintains the interface between the metallic and dielectric sections approximately parallel to the field direction. For the particles studied here, this means that two distinct motion regimes can be investigated, i.e. moving with a dielectric facing forward (ICEP) or a metal facing forward (sDEP).

To test whether the ICEP and sDEP electrokinetic propulsion mechanisms hold for the hemispherical-cone shaped particles, we investigated the active motion of the four different particles over range of frequences between 1 kHz and $10^3$ kHz. Figure 6 shows the superimposed bright field microscopy images of the observed particle motion perpendicular to the axis of the electric field for two different frequencies. Indeed, at a frequency of 10 kHz, particles moving with their dielectric parts forward can be observed (Figure 6a1-d1), while at 500 kHz, particles moving toward their metallic parts can be observed (figure 6a2-d2).

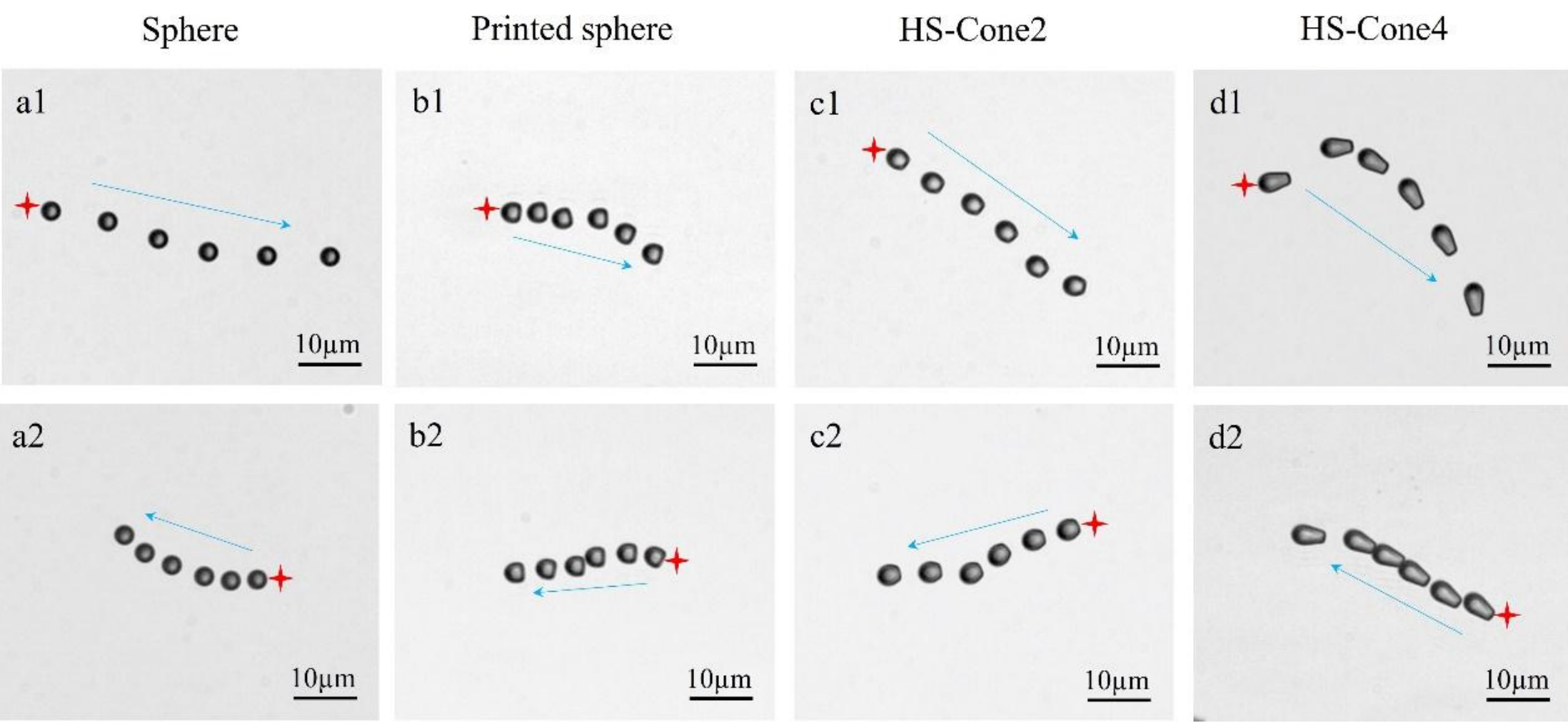

**Figure 6**. Superimposed bright-field micrographs of the four shapes of active particles migrating along the plane perpendicular to the electric field axis with different time steps. Red cross indicates the starting point and the blue line the direction of motion. The applied voltage was 6 $V_{pp}$ at a frequency of (a1-d1)10 kHz (forward dielectric parts) and (a2-d2) 500 kHz (forward metallic parts).

Clearly, the interplay between ICEP and sDEP also results in a distinct direction reversal for these particles. Another key observation is that the motion of all particles indeed occurred perpendicular to the field axis irrespective of applied frequency and thus HS-Cone2, HS-Cone4 and the printed sphere, all switch from moving forward with a flat sphere/conical protrusion to moving with the gold-coated hemispherical part forward.

To quantify the combined effect of the shape and the direction of motion on the propulsion velocity of the four different active particles, the MSDs were analyzed at different frequencies (see Figure S1, showing the MSD at 10 kHz, 100 kHz, and 500 kHz, respectively). Figure 7a shows the measured speeds for all frequencies from 1 kHz to 1 MHz, for the four different active particles. Note that the velocity is shown as determined with respect to the direction of the dielectric part, i.e. positive velocity for moving forward with their cone and negative velocity for moving forward with the gold-coated hemisphere. For all four particles shapes we observe a clear inversion of the velocity around a critical frequency $w_{cr} \approx 50$ kHz, which corresponds to the frequency expected for particles with a diameter of 3 μm and the low electrolyte concentration[16]. A slightly lower $w_{cr}$ is observed for HS-Cone4 that can be partly explained by an increase in volume but could also be caused by small deviations between different experiments. For the velocity at different frequencies for the four particle shapes, we observe several trends. For the Janus spheres the expected trend is observed, with relatively high speeds (~10 μm/s) in the ICEP regime ($< w_{cr}$) followed by a reduction in speed until half of the speed (~-5 μm/s) in recovered in the sDEP regime ($> w_{cr}$). For the small cone HS-Cone2 we find a similar trend and indicates the motion of these particles is dominated by the electrokinetic forces and is controlled by the field parameters. For the HS-Cone4 and the Printed Sphere, we observe that the velocities in both the ICEP and sDEP regime are similar (as highlighted by the plot of $v^2$ in Fig 7b). Interestingly, for HS-Cone4, the velocity in sDEP regime is even higher than for HS-Cone2, indicating that a larger conical protrusion increases the speed of the particles. Clearly, our observations show that even small geometrical changes have an impact on the propulsion of these active particles

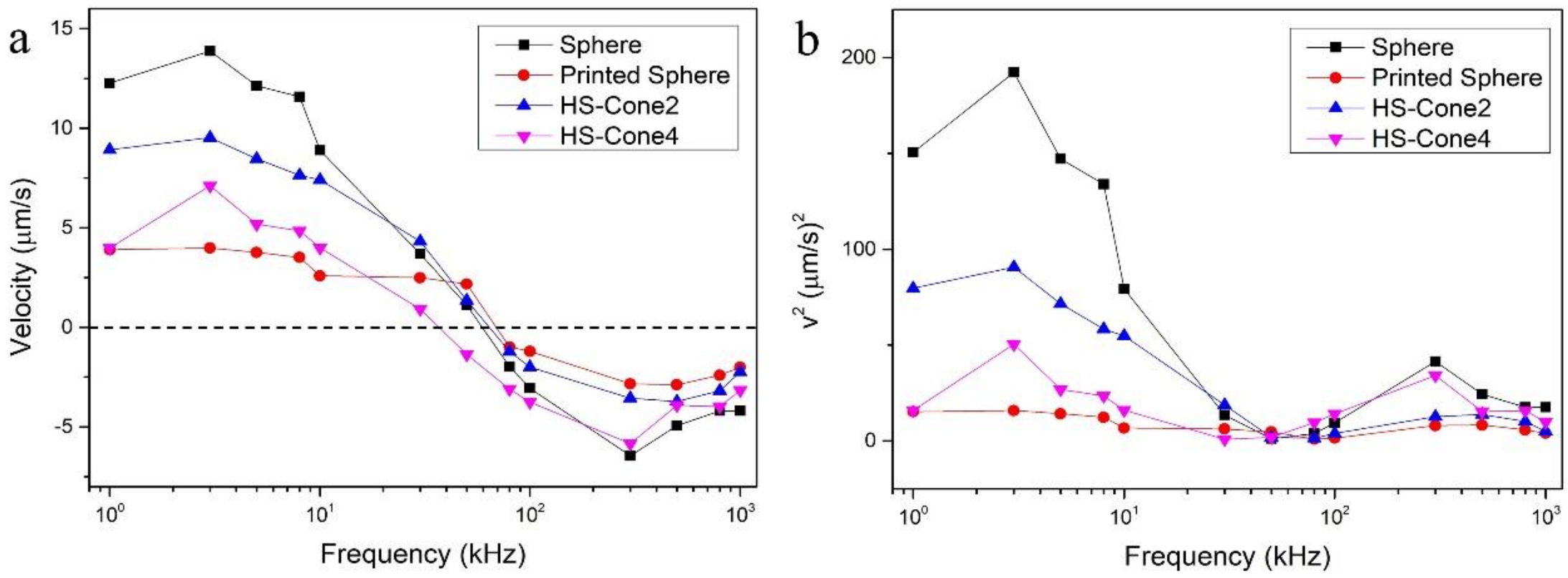

**Figure 7**. Frequency dependence of the self-propulsion velocity for different particles. The applied voltage is 6 $V_{pp}$.

To explain our observations for the different particle shapes, we consider two mechanisms that characterize active colloid motion. The first mechanism is the interplay between propulsion and random thermal fluctuations and the second mechanism is the self-generated fluid flows and the thickness of the hydrodynamic slip layer[34]. Firstly, considering the perfect Janus spheres, we expect random rotational motions will cause reorientation of the particle direction making the long-time dynamics Brownian and that the hydrodynamic slip layer can be considered constant since the surface properties do not change. Typically, however, out-of-plane rotations are not considered since these are hard to capture experimentally but also are not expected to influence the slip layer. Secondly, considering the cone-shaped particles we expect that out-of-plane rotations are suppressed due to their weight and forced alignment of their long axis to the wall. Consequently, we expect the hydrodynamic slip layer for the cone shape particles to be constant during propulsion. However, as we observe an increase in speed between forward and backward motion for HS-Cone4 with respect to HS-Cone2 and the spheres, we speculate that the hydrodynamic slip layer for the longer cones is changed upon direction reversal. Finally, for the printed spheres with a flat side we expect that gravitational alignment of the long axis is weaker, and that therefore out-of-plane rotations do occur. These rotations will change the alignment of the flat particle face with respect to the wall and can lead to changes in the particle motion, as observed for L-shaped colloids, where reflection of the particle orientation has been observed [18]. In addition, the change in particle orientation in turn might leads to a change in the hydrodynamic slip layer, where a reduction is expected to reduce the particle velocity[34].

## 4. Conclusion

In summary, we have studied the effect of a gradual shape change - from a sphere to a hemisphere with a conical protrusion – on the propulsion of anisotropic Janus colloids by combining 3D nano printing of anisotropic colloids, gold-coating and AC electric fields. Our quantitative analysis using video microscopy and particle tracking demonstrates that particle shape influences their behavior. We find that the addition of a conical protrusion to a hemispherical Janus particle systematically reduces its propulsion speed, with a longer cone resulting in a reduction in velocity compared to a simple sphere, while the presence of a flat face on a sphere seems to reduce the particle velocity drastically. In addition, we revealed that the fundamental electrokinetic mechanisms governing the particle motion in AC electric fields are robust to the studied shape changes and provide precise control. A distinct direction reversal observed at a critical frequency, governed by the interplay between ICEP and sDEP, was observed for all shapes. We found that shape does not influence the critical frequency for direction reversal but does affect the particle velocity, especially in the sDEP regimes where the large cone particles velocity increases with respect to the other shapes. Our results demonstrate that speed can be tuned not only by field parameters but also by morphological design.

By synergizing 3D printing with AC electric field propulsion, we have presented a robust framework for fabricating and controlling fuel-free anisotropic microswimmers that offers a reliable and remote-control mechanism provided by the interplay between ICEP and sDEP. We expect that this framework can provides clear guidelines for the design principles for engineering active colloids and will pave the way for creating sophisticated microrobots with tailored shapes for enhanced functionality in applications ranging from targeted delivery to complex micromanipulation.

## Acknowledgments

Z.Z. acknowledges financial support by Natural Science Research Project of Jiangsu Higher Education Institutions (no. 22KJA150008). J.M.M., D.K.M and T.W.J.V acknowledge financial support from the Netherlands Organization for Scientific Research (NWO) (Vidi.223.126). E.M.E.S. acknowledges financial support from HTSM.

**Author declarations section**

The authors declare no conflicts of interest. Author contribution: Conceptualization – ZZ, JMM; investigation – ZZ, DKM, EMES; formal analysis – ZZ, TWJV; methodology –ZZ, JMM; supervision – JMM; writing (original draft) – ZZ; writing (review & editing) – ZZ, JMM. All authors have read and approved the final text of the paper.

**Data availability statement**

Data for this article are available from the corresponding author on reasonable request.

# Effect of a Cone Shape on the Motion of Active Janus Colloids

# -

# Supplementary material section

Zhiyuan Zhao[1,2,3], Teun W.J. Verouden[2,3], Dillip K. Mohapatra[2,3], Evert M.E. Simons[2,3], Janne-Mieke Meijer[2,3,a)]

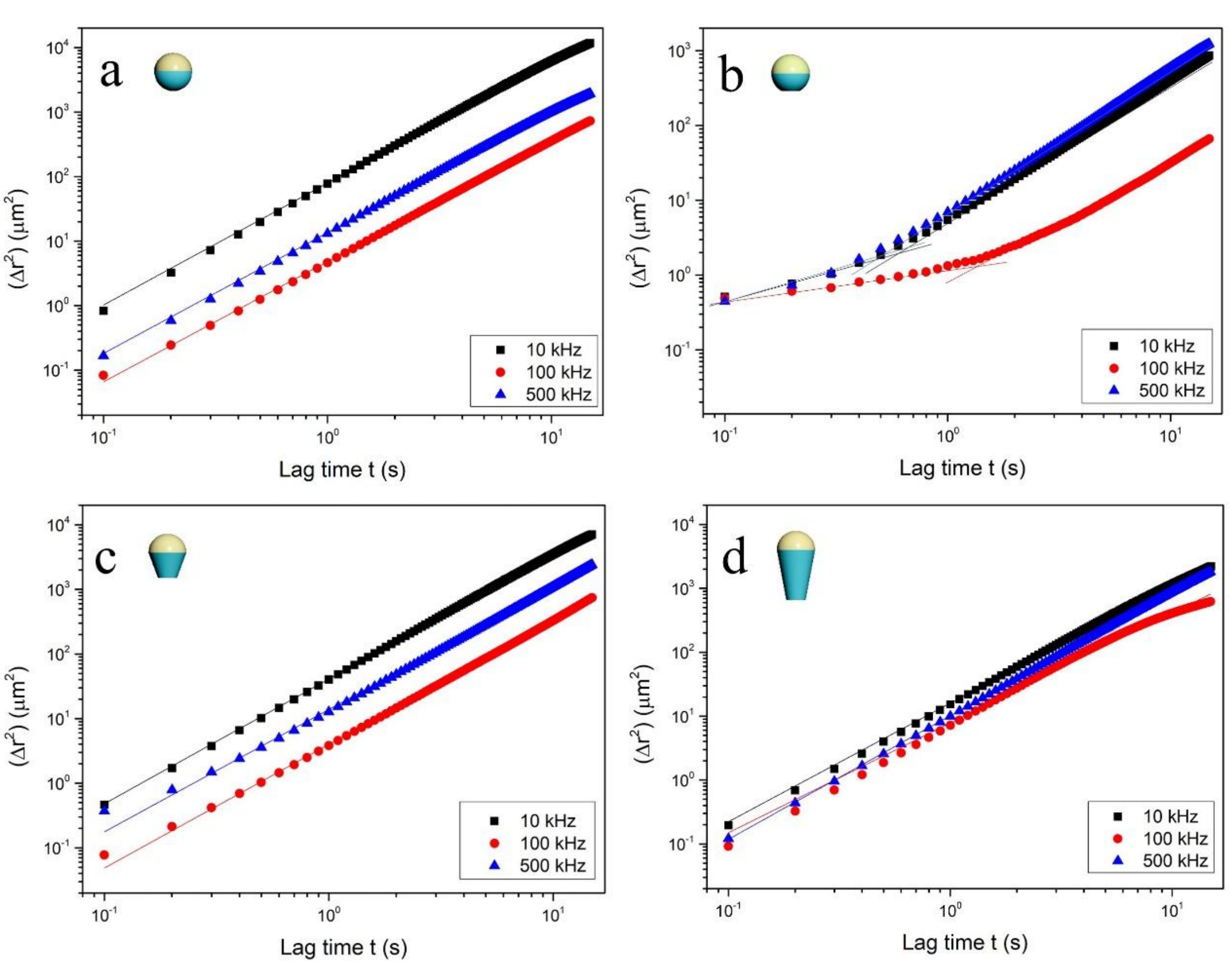

**Figure S1**. Mean Squared displacements of four shapes of active particles at 10 kHz, 100 kHz and 500 kHz are shown in log-log plot. (a) sphere, (b) printed sphere, (c) HS-Cone2, and (d) HS-Cone4. We note that the for the printed spheres (Fig 8b) some tracking errors occurred, leading to a plateau at short times.